# AN EXTENDED NETWORK CODING OPPORTUNITY DISCOVERY SCHEME IN WIRELESS NETWORKS

Yunlong Zhao[1,2], Zhao Dong[1], Masayuki Iwai[2],Kaoru Sezaki[2], Yoshito Tobe[3]

[1]School of Computer Science and Technology, Harbin Engineering University, 150001, Harbin, China
zhaoyunlong@hrbeu.edu.cn, dongzhao@hrbeu.edu.cn
[2] Center for Spatial Information Science, The University of Tokyo, 153-8505 Tokyo, Japan
masa@iis.u-tokyo.ac.jp, sezaki@iis.u-tokyo.ac.jp
[3] Department of Information Systems and Multimedia Design, Tokyo Denki University, 101-0054 Tokyo, Japan
yoshito_tobe@osoite.jp

## ABSTRACT

*Network coding is known as a promising approach to improve wireless network performance. How to discover the coding opportunity in relay nodes is really important for it. There are more coding chances, there are more times it can improve network throughput by network coding operation. In this paper, an extended network coding opportunity discovery scheme (ExCODE) is proposed, which is realized by appending the current node ID and all its 1-hop neighbors' IDs to the packet. ExCODE enables the next hop relay node to know which nodes else have already overheard the packet, so it can discover the potential coding opportunities as much as possible. ExCODE expands the region of discovering coding chance to n-hops, and have more opportunities to execute network coding operation in each relay node. At last, we implement ExCODE over the AODV protocol, and efficiency of the proposed mechanism is demonstrated with NS2 simulations, compared to the existing coding opportunity discovery scheme.*

## KEYWORDS

*network coding，coding opportunity，wireless network*

## 1. INTRODUCTION

Network coding (NC) is known as a new transmitting paradigm, which can efficiently increase the network capacity and throughput. In the traditional store and forward network, packets are only forwarded hop-by-hop by the relay nodes (intermediate nodes) along the routing path from a source to a destination, whereas the network coding can combine different packets into a new encoded packet then transmit it to next-hop node. Through the simple coding operation, network capacity and throughput can be improved greatly.

Figure 1 illustrates the basic idea of network coding in wireless networks. Figure 1(a) shows a simple network with chain topology, and Figure 1(b) shows a classic X topology network. In both cases, node C is a relay node for two flows. After receiving both $p$ and $q$, it can encode two packets and then broadcasts $p \odot q$ to two destinations. And the destinations can derive the expected native packet by decoding the encoded packet. As a result, the number of total transmissions is reduced from four to three in both scenarios. This is the basic idea of network coding, that it can improve network performance through reducing the number of transmissions and sending more information in a transmission by relay node.

When a relay node will encode two packets as an encoded one, there is a basic premise that it must guarantee the destinations can decode the native packet from the encoded one respectively. We only say that a node has a coding chance when this basic premise is met. Otherwise, the coding operation in a relay node will be meaningless. Apparently, How to determine that if there is a coding chance or not and how to discover all coding opportunities in each node is very important for network coding to improve network performance.

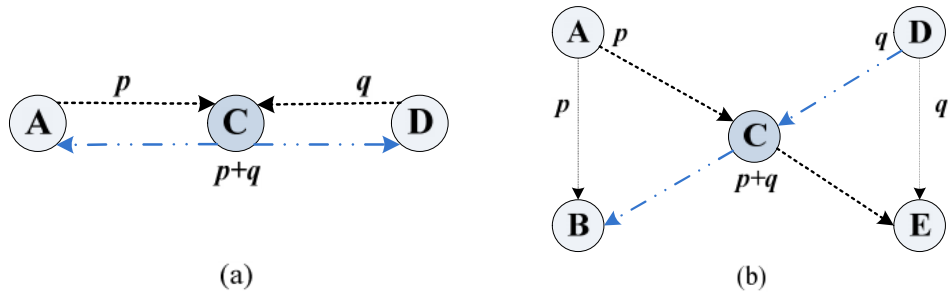

Figure1. examples of information transmission with network coding. Graph (a) is a simple chain topology, and graph (b) is a classic *X* topology.

Recently, opportunistic routing and coding-aware routing are regarded as an effective approach to create more coding opportunities actively. The key message is that the coding opportunity is taken into account when the routing decision is made rather than passively wait for the appearance of the coding opportunities. For instance, COPE[2] can discover XOR coding opportunities in practical network by the opportunistic listening and information exchange between one-hop neighbors. In COPE, each node announces to its neighbors that which packets it stores by sending reception reports each other, this can enable relay node detect coding chance actively. However, COPE has a fundamental limitation that it can only find coding chance within a two-hop region, i.e., the coding opportunities occur with in the region that only includes the relay node and its one-hop predecessors and successors. It is really a limitation for network coding to elevate network performance. Just like the figure 1, COPE can only find coding chance in specific networks, such as chain or X topology network. At present, how to explore more coding opportunities beyond the two-hop region is the essential issue.

Due to the broadcast nature of the wireless medium, when the source sends a packet to the destination, all its one-hop neighbors and neighbors of all relay nodes in the forwarding path can overhear the packet, so there must will be many copies of the packet in the network. We found that these copies are helpful to discover the coding chance in network. In this paper, we will propose an extended network coding opportunity discovery scheme ExCODE, which is suitable for wireless multi-hop networks, and can discover all coding opportunities in a node. Through appending additional information to a packet, ExCODE can expand the region of coding chance discovery to entire network. The additional information that is appended on each the packets is mainly about node IDs, which the packet has been sent or forwarded by them.

Before a packet is sent by a source node or forwarded by a relay node, the current node ID and its all one-hop neighbors' IDs should be appended to the packet. The appending information of the node IDs will let the next hop node knows that which other nodes have already stored the same packet. So when the current node receives another different packet also with the additional information that which nodes have a copy of the receiving packet, it can make a decision that whether there is a coding chance between these packets according to these information. This is the basic idea of ExCODE above. ExCODE can discover all coding opportunities in relay node through appending node IDs information on all packets.

In summary, this paper mainly contributes in the following three aspects.

• We develop ExCODE, which is universal and effective in detecting coding opportunities in wireless networks.
• Our scheme of coding opportunity discovery is also helpful for coding-aware routing protocol to select the best forwarder in a forwarder set.
• We evaluate the performance of ExCODE through simulation using ns-2, experimental results demonstrate the effectiveness of it.

The rest of this paper is organized as follows. In section 2, we describe the related works. And we describe the basic idea of ExCODE and design it in section 3. Simulation results are presented in Section 4. And then we make an analysis of the scheme in section 5. Finally, the conclusion is given in section 6.

## 2. RELATED WORK

Coding opportunity is very crucial to the practical coding scheme, which can utilize these opportunities to improve the throughput of multi-hop wireless network. Therefore, it is obvious that there are more coding chances there will be more times we can enhance the performance of network by executing coding operation. How to make full use of all available coding opportunities becomes a key issue. In [2], one of the first practical network coding schemes was proposed, named COPE, which uses deterministic information from the neighbor nodes to discover the coding opportunities. COPE is a new forwarding architecture for wireless network that inserts a coding shim between the IP and MAC layers, which identifies coding opportunities and benefits from them by forwarding multiple packets in a single transmission. The deterministic information is mainly about which particular packets have been stored in the neighbors, as a result, the relay node can try its best to discover the coding opportunities according to this information. In essence, COPE takes advantage of the 'broadcast nature' of the wireless channel to perform 'opportunistic overhearing' and 'encoded broadcast' so that the number of necessary transmissions can be reduced. However, COPE has a fundamental limitation that it can only find coding chance in two-hop region. Whether a node can execute coding operation crucially depend on traffic pattern. In other words, network coding is possible only when there exists certain 'coding structure' like 'X' or chain topology. Based on COPE, authors of [3] introduce the concept of coding-aware routing to actively maximize the overall coding opportunities in the whole network by using linear programming. Authors of [4] analyzed coding-aware routing, and regarded that it is an effective approach to actively create more coding opportunities. Its basic idea is that the coding opportunities are taken into account and are deliberately created during route discovery phase rather than passively waiting for the appearance of the coding opportunities. In [5], the authors propose a theoretical formulation for computing the throughput of network coding on any wireless network topology and pattern of concurrent unicast traffic sessions. In [6], the authors analyzed multiple scenarios and presented a generalized coding conditions based on the $k$-hop model coding structure. The conditions incorporate the fact that multiple flows may intersect with one flow at different nodes and guarantee all the flows involved in coding can be properly decoded at the destinations. In [7], the authors formally establish coding conditions for a very general scenario that multiple coding nodes may exist along a path and multiple flows may intersect at one node. they systematically analyze possible coding scenarios and develop generalized coding conditions to ensure the decoding ability at the destinations. And they propose a novel coding-aware routing metric, the free-ride-oriented routing metric (FORM), with the objective of exploiting the coding opportunities so that a new flow can free ride on the existing traffic and can be supported with a smaller number of transmissions. In [8], the first distributed coding-aware routing system for wireless networks was proposed, named DCAR. It is an on-demand and link-state routing protocol, and it incorporates potential coding opportunities into route selection using the "Coding + Routing Discovery" and "Coding-aware Routing Metric" (CRM). DCAR also adopts a more generalized coding scheme by eliminating the "two-hop" limitation in COPE. In [9], the authors develop a theoretical formulation for computing the throughput of network coding on any wireless network topology and any pattern of concurrent unicast traffic sessions, and provide a systematic method to quantify the benefits of using network coding in the wireless networks. In [10], the authors stated other limitation of COPE, whose implementation only fits in LS routing protocol, and then proposed an improved scheme to discover coding opportunities, which fits in not only link-state routing protocols but also distance-vector protocols. In [11], the authors proposed a complex optimization framework for adaptive coding and scheduling. In [12], the authors studied limitations of COPE under practical physical layer and link-scheduling algorithms, proposed the concept of coding-efficient link scheduling for practical network coding. In [13], the authors proposed a Backbone Routing with Network Coding (BR-ONC) scheme over multi-hop wireless network, which combines the benefits of both backbone routing and network coding techniques. With backbone-based routing, all packets are forced to be transmitted over a constructed backbone. In [14], the authors present a Probabilistic Network Coding with Priority (PNCP) method, it keeps different queues in node buffer for different flows. Whether a packet is transmitted with or without network coding is determined by its priority and the queue state. In [15], the authors proposed a real-time and high coding opportunity discovery scheme. All the aforementioned work belong to proactive mechanism with the lack of generality and flexibility, because some deterministic information must be constructed before the relay node makes decision to encode the packets or not. In this paper, we propose an extended network coding opportunity discovery scheme, named ExCODE. Compared with former work, ExCODE is the first real-time, universal, distributed and practical coding-aware routing system, and can exploit more coding opportunities with less transmitting overhead.

## 3. ExCODE

We introduce ExCODE, a new network coding opportunity discovery scheme for multi-hop wireless network. It takes advantage of the broadcast characteristic of a wireless channel for data delivery, appends extra information on all of packets that is forwarded or sent firstly by a node. The information is mainly about ID node and all of its one-hop neighbors IDs. This operation can enables ExCODE to detect more coding opportunities and exploits them to forward multiple packets in a single transmission. For ease of description, we first define the following few terminologies:

- **Native packet**: A primary packet that is sent by a node firstly
- **Encoded packet**: A packet that is the XOR of multiple native packets
- **Coding node**: A node which encodes packets, e.g., node C in Figure 1
- **Node ID**: A 32-bits IP address, identifies a node uniquely in the network
- **IDs set**: includes all IDs of every neighbor of a node

- **Set$_p$**: IDs set that is appended on the packet $p$
- **Sender**: A node that sends a packet initially
- **Forwarder**: The next-hop node selected for continuing to transmit a packet
- **Input queue**: A FIFO queue at each node, where it keeps the packets it has received or overheard
- **Output queue**: A FIFO queue at each node, where it keeps the packets it needs to forward

### 3.1. A Simple Case Study

It is evident that network coding is very helpful to improve network performance, so it is much better to exploit more coding opportunities and make use of it in real time. But which different packets can be encoded together or not at current relay node is a key issue, it also means how to discover the coding opportunities at current relay node. Let's begin with a simple case to explore the coding opportunities discovery scheme.

Figure 2(a) shows a very simple network topology, suppose node A will send packet $p$ to G along the path A−>C−>E−>G, and node F will send packet $q$ to node B along the path F−>D−>C−>B. According to the COPE method, when the intermediate node C receives $p$ and $q$ one after another, it will not encode them with $p \oplus q$. Because it doesn't know whether node G has already stored packet $q$ or not, and dose not guarantee the destination E can decode the native packet $p$ from the coding packet $p \oplus q$ successfully. Similarly, node C also can not guarantee the destination B to succeed decoding. Consequently, node C will not take network coding, which means it will miss a coding chance. In fact, when node F sends packet $q$ to node B, node G can overhear it, because node G is one-hop neighbor of node F. So node G can decode its native packet $p$ from the coded packet $p \oplus q$. Similarly, node B also can decode native packet $q$ from $p \oplus q$.

COPE only exchange reception report between one-hop neighbors, this lead to it can merely discover coding chance in two-hop region. If we can extend the region to multi-hops, the coding opportunity will not be missed in figure1. Considering above case, if the relay node C can know such information that which nodes have received the packets $p$ and $q$, it can discover all available coding opportunities to avoid missing coding chance.

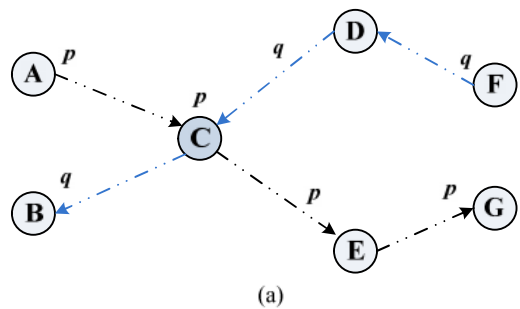

(a)

| packet | packet holders |
|--------|----------------|
| *p* | A、B、C |

(b)

| packet | packet holders |
|--------|----------------|
| *q* | F、G、D |

(c)

Figure2. A simple coding case. Graph (a) is a basic network model. The information that is appended on packet *p* is described in table (b), includes all node IDs that which nodes can overhear the packet before it arriving the current node.. The information that is appended on packet *q* is described in table (c).

Our solution is provided as follows. When the source, node A, initiates packet *p*, it will append itself node ID and all its neighbors' IDs, i.e. node B and C, to packet *p*, then sends it out. After node C receives packet *p*, it can know that both node B and A have stored the copy of packet *p* in their own buffers. Similarly, node C can also know that both node D and G have stored the copy of packet *q* in their own buffers. By simply analysis and comparison, node C can easily discover the coding opportunities.

## 3.2. Basic Idea

When a relay node receives two packets from two different traffic flows, it maybe has the chance to encode them, and then it will activate the coding opportunity discovery procedure. COPE can only discover the coding opportunity within two-hop region away from the current relay node, which also rely on the whole network topology information. The method presented in [6] relies on the two-hop neighbor information and sender information to discover the coding chance. In fact, it is enough for the relay node to discover all the coding opportunities if it knows such information that which nodes have already stored the packets it have received, no matter through the normal forwarding them or overhearing them.

Basing on the study and analysis of above simple case, a new coding opportunity discovery scheme will be proposed, named extended network coding opportunity discovery scheme (ExCODE). Due to the broadcast nature of the wireless medium, when a node sends a packet to another, all its one-hop neighbors and neighbors of all relay nodes in the forwarding path can overhear the packet, so there must will be many copies of the packet in the network. We will use the information of all these copies of packet to find coding chance.

We describe ExCODE as follows: For any packet, before it is sent by a source node or forwarded by a relay node, the current node ID and its all one-hop neighbors' IDs should be appended to the packet. The appending information of the node IDs will let the next hop node, e.g. relay node, knows that which other nodes have already stored the same packet. Such appending information is called the *packet holders*, just like table (b) and table(c) in figure 2. It is a node set that all have overheard the packet, if the packet is *p*, we will call the set is $Set_p$. So when the current relay node receives another different packet also with the additional information that which nodes have a copy of the receiving packet, it can make a decision whether to encode the two packets or not according to the following rules:

a) The first receiving packet, named $p$, has already been stored in such nodes which is the destination node of another receiving packet, named $q$. This condition can be easily determined by simply comparing the $q$'s destination ID with the $p$'s holders IDs $Set_p$.

b) At the same time, the second condition should also be satisfied, which $q$ has already been stored in such node which is the destination node of $p$. Such condition can also be easily determined by simply comparing the $p$'s destination ID with the $q$'s holders' IDs, e.g. $Set_q$.

Assume there is a packet $p$, its destination is node $D_1$, and there is another packet $q$, its destination is node $D_2$. The IDs information is appended on $p$ is $Set_p$ , appended on $q$ is $Set_p$.We can describe these rules above by a formula:

$$D_1 \in Set_q \,\&\&\, D_2 \in Set_p \qquad (1)$$

If there are two packets $p$ and $q$ in a relay node can enable the consequence of the formula to true, these packets must can be encoded together and the destinations must can decode the native packet from the encoded packet. The detailed description of ExCODE is illustrated in next section, and it can be extensively used in multi-hop wireless network.

## 3.3. ExCODE Design

Before giving the formal description of ExCODE, we should make some assumptions. Firstly, there is an *input queue*, an *output queue* and a *buffer* in each node to receive, forward and temporarily store the packets, and each capacity is unlimited. Secondly, we assume that the wireless link is reliable and symmetric, which means each neighbor nodes pair can receive and overhear the packets with each other. Lastly, the transmitting packets have enough reserved space to accept the additional information.

ExCODE accompanys the data flow and is implemented by each node along the flow, which is activated when a node receives a new packet according to the two sub-procedures described in Algorithm 1 and Algorithm 2 respectively. Every time a node will send a packet to another, it appends additional information on it firstly. We assume the relay node is O, the sending procedure is depicted in Algorithm 1:

---
Algorithm 1 Sending Procedure
---

While (the *output queue* is not empty) {
      Node O picks a packet $p$ from the head of *output queue*;
      Node O appends its ID and its one-hop neighbors' ID to $p$;
      Forward $p$ out;
      } //end of While
Return; //The end of Sending Procedure

---

When a packet arrived in relay node O, it should be put in *input queue* firstly. And then the node picks the packets from the *input queue* sequentially and activate the procedure of coding chance discovery, detects that if there is a opportunity that the packet can encoding with other packet. The receiving procedure is depicted in Algorithm 2:

Algorithm 2 Receiving Procedure

While (the *input queue* is not empty) {
        Node O picks a packet $p$ from the head of *input queue*;
    If( the node has received $p$ before){
      Discard the packet $p$;
      exit(0); }
   If ($p$ is just an overheard packet) {
     store $p$ to the *buffer*;
     exit(0); }
   else if ( $p$'s destination node is node O) {
        if ($p$ is a encoded packet)
          Node O decodes it and then receives it;
        else Node O receives it directly;
         exit(0); }
    else {
        for (i=1; i<n; i++) { // Assume there are n packets in *input queue*.
         if ($p$'s destination ID is in the $p_i$'s holders set && $p_i$'s destination ID is in the $p$'s
holders set) {
                store $p$, $p_i$ to the *buffer*;
                $p_{en} = p \oplus p_i$ ;
                put $p_{en}$ in the end of *output queue*;
                store $p_{en}$ to the *buffer*;
                remove $p$, $p_i$ from *input queue*;
                exit(0); }
          else{
            put $p$ in the end of *output queue*;
            store $p$ to the *buffer*;
            remove $p$ from *input queue*;
           }
         }
      }
    }//end of While
Return;    //The end of Receiving Procedure

## 3.4. An Enhanced Case Study

In section 2.1, we have demonstrated that the proposed scheme can increase the coding opportunities in a simple example. Next, we will expand the example to general situation. Furthermore, the ExCODE scheme is also effective in the more complex multi-hop wireless network scenarios.

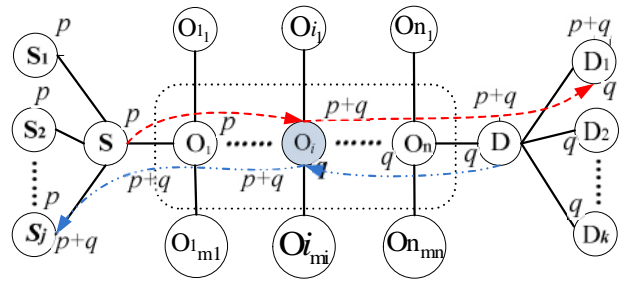

(a)

| packet | packet holders |
|--------|----------------|
| $p$ | S、$S_1$、$S_2$···$S_j$、$O_1$、$O_{1_1}$···$O_{1_{m1}}$、$O_2$、$O_3$···$O_{i-1}$ |

(b)

| packet | packet holders |
|--------|----------------|
| $q$ | D、$D_1$、$D_2$···$D_k$、$O_n$、$O_{n_1}$···$O_{n_r}$、$O_{n-1}$···$O_{i+1}$ |

(c)

Figure 3. An enhanced coding case. Graph (a) is a basic network model. The information that is appended on packet $p$ is described in table (b), includes all node IDs that which nodes can overhear the packet before it arriving the current node. The information that is appended on packet $q$ is described in table (c).

In Figure 3, we extend the sample topology in figure 2 to a more complex situation. There are more nodes here, and each node has several neighbors. Suppose that nodes $O_1$, $O_2$ ... $O_i$ ...and $O_n$ are relay nodes, and each node has many one-hop neighbors, e.g. node S has $j$ neighbors $S_1$, $S_2$…$S_j$, node $O_1$ has $m_1$ neighbors $O_{11}$…$O_{1m1}$, node D has $k$ neighbors $D_1$, $D_2$…$D_k$, and $j$, $m_1$, $m_2$…$m_i$…$m_n$ and $k$ are not equal. The dashed rectangular box refers to that there are many nodes between $O_1$, $O_i$ and $O_n$, and there are also many others nodes which are the one-hop neighbors of nodes $O_1$, $O_2$…$O_i$…and $O_n$.

Assume that node S will send packet $p$ to node $D_1$ along the path S−>$O_1$... $O_i$ ...−> $O_n$−>D −>$D_1$, and node D will send packet $q$ to $S_j$ along the path D−>$O_n$ ...$O_i$ ... −>$O_1$−>S−>$S_j$, as shown in dashed line arrow. Table (b) records the node IDs that which nodes can overhear the packet before it arriving the current node. According to the COPE method, when the relay node $O_i$ receives $p$ and $q$ one after another, it will not encode them with $p \odot q$, because there is no enough information enables the node $O_i$ to know that two destinations $D_1$ and $S_j$ can extract native packet respectively.

In ExCODE, we append additional node IDs information on all packets, it enables every relay node to know that which nodes have received or overheard the packet before. Figure 3(b) and (c) shows the information that is carried by packet $p$ and $q$. We can see from the Figure 2 (b) and (c) that node S, $S_1$ and many others nodes hold a copy of packet $p$, nodes D, $D_1$ and many other nodes hold a copy of packet $q$. The packet was forwarded along a routing path, all relay nodes and their 1-hop neighbors can receive or overhear it. For the broadcast nature of wireless medium, thus all of them hold a copy of the packet. ExCODE enforce all of relay nodes appending itself node ID and its one-hop neighbors' IDs to current packet. Through this operation, each relay node can know which nodes has a copy of the packet, thus discover more coding chance.

In this example, through checking their *packet holders* information, node $O_i$ can know that node $S_j$ has stored $p$, and node $D_1$ has stored $q$. These packets $p$, $q$ and IDs sets $Set_p$, $Set_q$ meet the formula (1), namely the value of "$D_1 \in Set_q$ && $S_j \in Set_p$" is true here. So, it can encode these packets and then forward the encoded packet $p \odot q$ to the destinations, $S_j$ and $D_1$ can extract the native packet easily. It is obvious that ExCODE is really a useful scheme to discover coding opportunities.

In a word, so long as two packets which belong to two different flows, are received by any relay node at the same time, the relay node can check their *packet holders* information, if each packet's destination can be found in another *packet holders* set, there must be a coding chance. Through adding node IDs to the packets, this scheme can finds all coding chances in relay nodes.

## 4. SIMULATION AND RESULTS

In this section we evaluate the performance of ExCODE through simulation experiments using ns-2, and compared ExCODE with COPE and non-coding system. We deployed 16 static nodes in an $800 \times 800$ m$^2$ square field. The radio range is approximately 200 meters. We randomly generated UDP flows and varied the offered load by adjusting the number of flows. All the flows are constant-bit-rate(CBR) flows with fixed packet size 512 bytes. We used 802.11b as the underlying MAC-layer protocol. The channel rate was set to be 2 Mb/s. We used AODV as the routing protocol, the source and the destination of each flow were randomly chosen. Each simulation lasts for 120 seconds.

The result of the throughput comparison is shown in Figure 4 where ExCODE, COPE, and non-coding system is compared respectively.

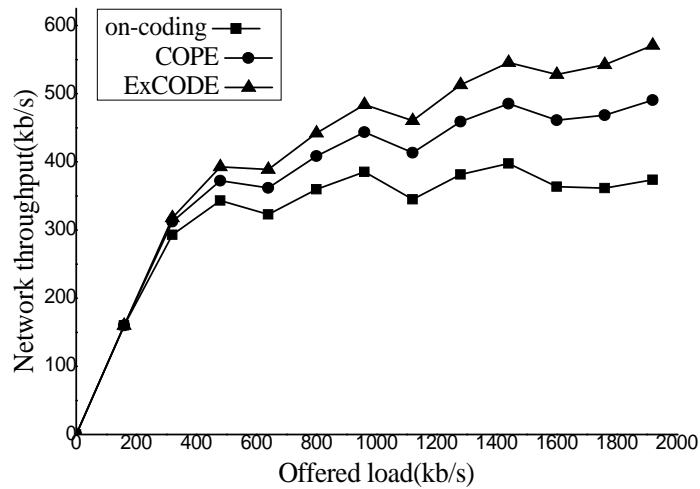

Figure 4. Network throughput vs. offered load

It is shown that ExCODE can improve the network throughput significantly as compared with COPE and non-coding system under different offered loads. When the offered load is small, there are not many packets arrive in the nodes in a period of time; it is difficult to find inter-flow coding chance for relay nodes. In this time, the throughput of ExCODE, COPE and non-coding system is approximately similar. As the offered load increases, the probability of several packets meeting in a node is larger in a moment. It is convenient for relay nodes to find coding chance, both ExCODE and COPE can obtain more network throughput than non-coding system. At the same time, because ExCODE can discover more coding chances than COPE, so the former can gain more network throughput than the latter. On average, ExCODE achieves 13 percent throughput gain over COPE and non-coding system with a maximum gain of 16 percent (over COPE) and 53 percent (non-coding system), respectively. As a result, the throughput of ExCODE is larger than COPE and non-coding system.

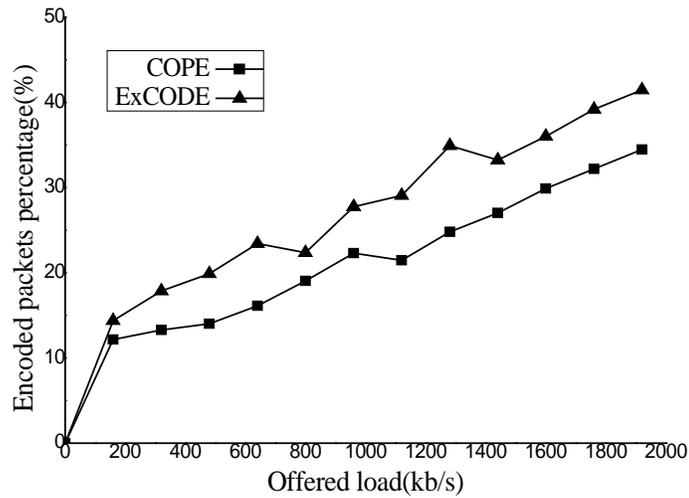

Figure 5. Encoded packets vs. offered load

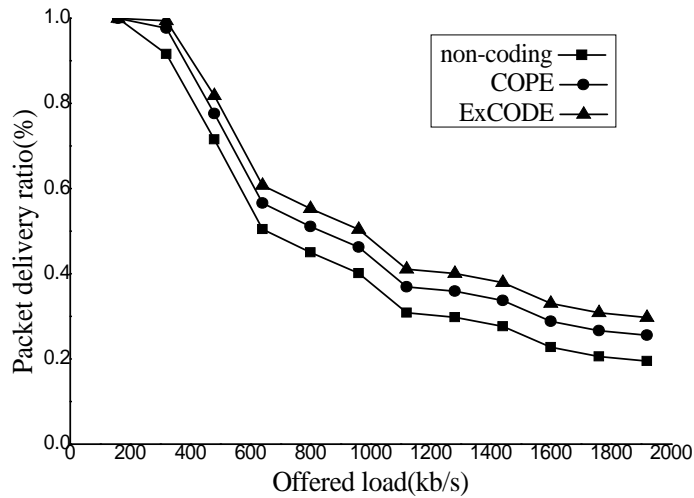

Figure 6. Packet delivery ratio vs. offered load

Figure 5 plots the percentage of encoded packets with ExCODE and COPE respectively under different offered loads. It is shown that ExCODE can take more primary packets to be encoded packet as compared with COPE under the same offered loads, and the former always has more encoded packet than the latter with the increasing of the offered load. It means that ExCODE can really discover more coding chance than COPE at the same situation. On average, ExCODE always can discover more 10 percent coding chance than COPE.

Figure 6 shows the packet delivery ratio with ExCODE, COPE, and non-coding system under different offered loads respectively. With the increasing of the offered load, the packet delivery ratio of all ExCODE, COPE and non-coding system are smaller than before, but ExCODE can gain more percentage than COPE and non-coding system from the beginning to the end. It is obvious that ExCODE can gain more coding chances and then uses them to increase the packet delivery ratio.

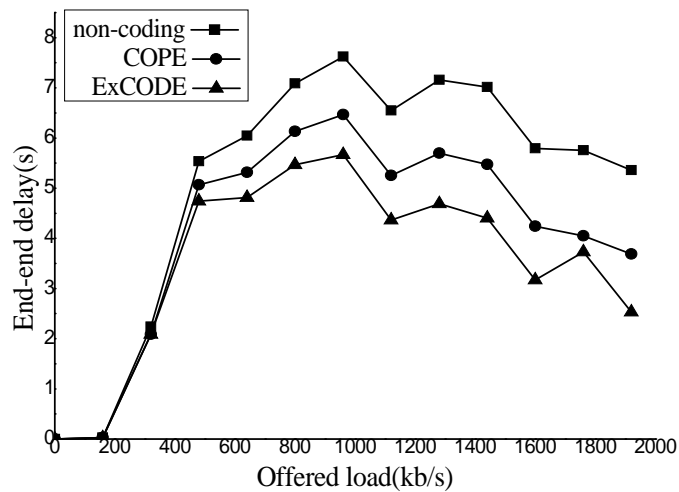

Figure 7.  End-end delay vs. offered load

The result of the end-end delay comparison is shown in Figure 7 where ExCODE, COPE, and non-coding system is compared under different offered loads respectively. It is shown that ExCODE has smaller end-end delay as compared with COPE and non-coding system. Because ExCODE can discover more coding chances to encode packets, and then forwarding more packets in a period of time. As a result, it can decrease the time to forward the same amount of packets.

## 5. ANALYSIS OF EXCODE

Through appending the *packet holders* information to the current packet, ExCODE indeed can find more coding chance in the relay nodes. Whenever a packet arrives at a relay node, the node would know immediately whether there is another packet in the *input queue* can encode with it, and judge that whether there is a coding chance or not through checking their *packet holders* information. It is really a high coding chance discovery scheme.

We have compared the scheme with COPE which presents two methods to discover the XOR coding opportunities. The first one is depending on reception report, each node broadcast its reception report to tell its neighbors that which packets it has stored. Through reception report, COPE can find the coding chance within a two-hops region. At the meanwhile, a node can encode the packets which both their source and destination are neighbors of the current node. As the second method, COPE guesses whether a neighbor has a particular packet intelligently. It would be used while the node did not obtain the reception report. It is same as the first method, and the coding structure is just limited within a two-hop region.

Our scheme of coding opportunity discovery is also helpful for coding-aware routing protocol to select the best forwarder in a forwarder set. It is the most important for coding-aware routing protocol to select a favorable next-hop node to forward the packet. In this paper, we have prove that ExCODE can discover all coding chance in relay nodes, so we can make use of it in computing of coding opportunities for each node in certain forwarder set, and select the best next-hop forwarder according to the result of computing before. We confirm that the computing result is accurate and really useful for coding-aware routing protocol to select the best forwarder in all next-hop nodes.

From above discussion, we know that COPE can not discover the coding chance beyond two-hops region. Compared with COPE, ExCODE is more efficient. So long as the destinations nodes of two packets can be found in another *packet holders* set, there must be a chance to encode them in current node. Namely, we expand the coding region to the n-hops. In conclusion, ExCODE has some benefits as follows:

Firstly, it is a real-time and high coding opportunity discovery scheme. Secondly, it is independent on any kinds of route protocols. Lastly, it extends the region of discovering coding chance to n-hops.

## 6. CONCLUSIONS

ExCODE is proposed in this paper, which is an extended network coding opportunity discovery scheme. It can effectively extend the region of coding discovery to the n-hops, and can exploit more coding opportunities. Through the comparison and analysis between ExCODE and COPE, ExCODE can indeed discover more coding chance than COPE, and can be applied in any kind of wireless routing protocols to enhance the coding-aware function. Our future work is about to practice ExCODE in a real network to test its performance, such as a wireless sensor network that is built up by SunSpot sensors.

## ACKNOWLEDGEMENTS

This research was supported by the National Natural Science Foundation of China under Grant No. 61003235, the Natural Science Foundation of Heilongjiang Province of China under Grant No. F200902, the Fundamental Research Funds for the Central Universities under Grant No. HEUCF100607, the Educational Commission of Heilongjiang Province of China under Grant No. 11553047 and Harbin Scientific & Technological Innovation Research Funds under Grant No. C2011QN010005.

## Authors

Yunlong Zhao, associate professor, received his B. S. degree in computer science and engineering, M. S. and Ph. D. degrees in Computer Architecture from the School of Computer Science and Technology, Harbin Institute of Technology , China , in 1998, 2000, and 2005, respectively. His research interests include wireless network, network coding, pervasive computing and Internet of things, etc. Dr. Yunlong Zhao has acted as reviewer for several prestigious journals, such as Journal of Communications and Networks, International Journal of Information Technology, Journal of Computer Applications, and so on. He was a recipient of many Awards during studying in Harbin Institute of Technology. He has published more than 40 papers in public, and half of them have been indexed by SCI, EI and IEEE member, CCF senior member and CCF computer architecture society fellow.

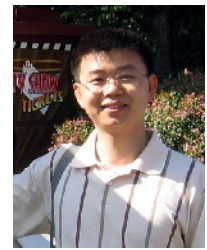

ISTP. Dr. Zhao is an

Zhao Dong is a Master in College of the Computer Science and Technology, Harbin Engineering University. His research interests include wireless network, network coding, and security issues on the Internet.

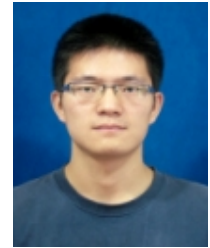